\documentclass{article}
\usepackage{ijcai20}

\usepackage{times}
\usepackage{soul}
\usepackage{url}
\usepackage[hidelinks]{hyperref}
\usepackage[utf8]{inputenc}
\usepackage[small]{caption}
\usepackage{graphicx}
\usepackage{amsmath}
\usepackage{amsthm}
\usepackage{booktabs}
\usepackage{algorithm}
\usepackage{algorithmic}
\usepackage{amssymb}
\usepackage{units}
\usepackage{xcolor}
\usepackage{xargs}    
\usepackage[export]{adjustbox}

\newcommand{\e}{\varepsilon}

\newcommandx{\rlm}[2][1=]{\todo[linecolor=violet,backgroundcolor=violet!25,bordercolor=violet,#1]{\textbf{Romain:} #2}}
\newcommandx{\rtcm}[2][1=]{\todo[linecolor=red,backgroundcolor=red!25,bordercolor=red,#1]{\textbf{Remi:} #2}}

\title{Reinforcement Learning Framework for Deep Brain Stimulation Study}

\author{
Dmitrii Krylov$^1$\and
Remi Tachet$^2$\and
Romain Laroche$^2$\and
Michael Rosenblum$^3$\And
Dmitry V. Dylov$^1$
\affiliations
$^1$Skolkovo Institute of Science and Technology, Bolshoy blvd. 30/1, Moscow, 121205, Russia\\
$^2$Microsoft Research Lab, 550-2000 McGill College Ave, Montr\'eal H3A 3H3, Canada\\
$^3$
University of Potsdam, Karl-Liebknecht-Str. 24/25, 14476 Potsdam-Golm, Germany
\emails
\{Remi.Tachet, Romain.Laroche\}@microsoft.com,
mros@uni-potsdam.de,
d.dylov@skoltech.ru
}

\begin{document}

\maketitle

\begin{abstract}
Malfunctioning neurons in the brain sometimes operate synchronously, reportedly causing many neurological diseases, e.g. Parkinson’s. Suppression and control of this collective synchronous activity is therefore of great importance for neuroscience, and can only rely on limited engineering trials due to the need to experiment with live human brains. We present the first Reinforcement Learning (RL) gym framework that emulates this collective behavior of neurons and allows to find suppression parameters for the environment of synthetic degenerate models of neurons. We successfully suppress synchrony via RL for three pathological signaling regimes, characterize the framework’s stability to noise, and further remove the unwanted oscillations by engaging multiple PPO agents.
\end{abstract}

\section{Introduction}
A hypothesis in neuroscience claims that several neurological diseases, such as Parkinson's\footnote{Parkinson's disease is the second most common neurodegenerative disorder after Alzheimer's. It affects approximately seven million people globally and 1–2 per 1000 of the population at any time. Its prevalence is increasing with age affecting 1\% of the population above 60 years~\cite{Tysnes2017}.}, originate from the networks of pathologically synchronous neurons in the brain. These malicious ensembles of neurons can collectively generate signals in a synchronized manner, debatably leading to the ``macro'' symptoms such as tremor, rigidity, bradykinesia, postural instability, and other movement abnormalities~\cite{Johnson2008,Gradinaru-09,Deniau_et_al-10}. To overcome these collective signals (or 'modes') in advanced stages of a disease, doctors often resort to high-frequency open-loop pulse stimulation of certain brain regions via implanted micro-electrodes --  a technology called deep brain stimulation (DBS)~\cite{Benabid_et_al-91,Benabid_et_al-09,Kuehn-Volkmann-17}.

Today, DBS systems have no feedback algorithms embedded into their circuitry, with doctors simply adjusting the electrode currents according to the symptomatic observations~\cite{Kuehn-Volkmann-17}. Although the new generations of DBS promise to provide the feedback functionality, the difficulty of conducting experimentation with live human brains still makes it hard to find the best stimulation algorithm experimentally. Moreover, a large network of interacting neurons is a complex non-linear system, which, considering limitations of the hardware and the unknown biological pathway of the illness itself, calls for additional modeling effort. 

As such, there appeared a demand for synthetic physical modeling to mimic the collective signaling patterns of neuronal ensembles~\cite{Hansel92,GIELEN2001ix,GOLOMB}. The aim of several open-loop~\cite{Tass-01} and of the more recent closed-loop feedback-based control approaches~\cite{Rosenblum-Pikovsky-04,Rosenblum-Pikovsky-04a,Popovych-Hauptmann-Tass-05,Lin_2013} is to desynchronize the large network of neurons, without suppressing the very oscillatory activity of individual neurons. In such physical synthetic models, the output of neurons is typically described either by several sets of ordinary differential equations (ODE), by partial differential equations (PDE), or by a map-based definition.

At the same time, the explosive development of RL~\cite{RL-Sutton1998} in recent years has offered a new data-driven methodology that could operate completely unaware of the physical world or of the underlying neuronal model. The Machine Learning (ML) techniques are now extensively used for analysis and prediction of complex systems~\cite{Parlitz-18,PhysRevLett.120.024102,Zimmermann-Parlitz-18,Quade-18,Cestnik-Abel-19,PhysRevE.99.042203,Yeo-Melnyk-19} and it seems natural to propose this framework for the purposes of control in deep brain stimulation as well. RL is often difficult to apply to real-world applications because of the necessary exploration, which implies a large number of trial and errors, potentially with dramatic consequences, before being able to improve the policy. Nevertheless, DBS is a setting where those drawbacks are absent. Its action space can easily be constrained to ensure that the agent's actions are harmless to the patient, and, depending on the DBS device, the frequency of decision making ranges from 60 Hz to 150 kHz~\cite{Su2018}, meaning that 1 million transitions may be collected in less than 2 to 5 hours on a single patient. 

\begin{figure}[h!]
\centering
\includegraphics[width=1\columnwidth]{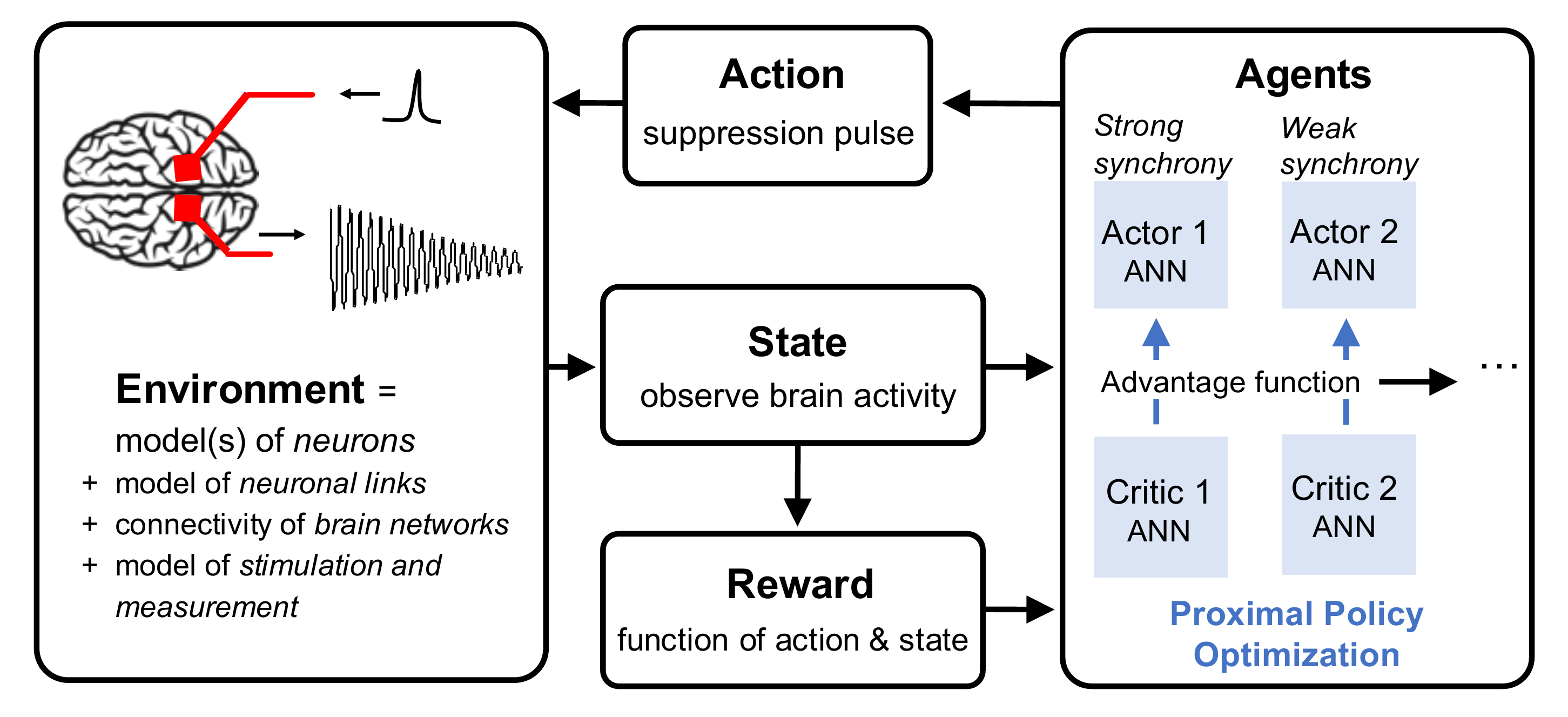}
\caption{Concept of the proposed framework for modeling interaction with a series of different models of neuronal ensembles via RL. Due to the highly nonlinear nature of the environments, multiple RL Agents can be used for different strengths of synchrony to achieve finer control.
}
\label{fig:diagram}
\end{figure}

In this paper, we report creation of a convenient gym environment~\cite{brockman2016openai} for developing and comparing the interaction of RL agents with several types of neuronal models developed in computational neuroscience and physics. 
The ODEs or descriptor maps are wrapped into the framework as individual environments, allowing to switch easily between environments, to use various RL models, and potentially multiple agents. Using this framework, we demonstrate successful suppression of the collective mode in three different types of oscillatory ensembles, using various policy-based approaches~\cite{RL-Sutton1998}, and show the first demonstration of synchrony suppression using a pair of RL agents trained using PPO. The suppression workflow proposed here is universal and could be used to create benchmarks among different physical models, to create different control algorithms, and to pave the way towards the clinical realization of deep brain stimulation via RL. The policy gradient algorithm PPO used below can provide a robust data-driven control, agnostic of the neuronal model and promises pathways for integration with current clinical DBS systems.

\section{The model}
In this work, we train RL agents with proximal policy optimization~\cite[PPO]{RL-PPO-schulman2017} (see diagram of Fig.~\ref{fig:diagram}). Classically, training involves five main blocks for the control problem: Environment, Action, State, Reward, and Agent. The flow works as follows: the agent observes a state, then takes an action, next, the environment responds with a reward signal and the agent observes the new state of the environment, which closes the loop of interaction. We now describe each block, its characteristics, and its function in detail.

\subsection{Environment}
Fig.~\ref{fig:diagram} conceptually shows which components contribute to the model of our RL ``environment''. Each configuration, such as the model and the number of neurons in an inter-connected ensemble, type of their links, strength and the model of connectivity within the ``brain'', can be tuned to simulate certain pathological signalling patterns. Well studied in the physical sciences, such models of pathological brain networks include (ranging from simple to complex):
a globally coupled ensemble, interacting groups of excitatory and inhibitory neurons, including spatially-structured ones, detailed models of involved brain regions, and other more complex models. 

Within these models, individual neurons could be described by (from simple to complex): map-based models (e.g. Rulkov), integrate-and-fire models, conductance-based models (simple 2D models of spiking dynamics, e.g. Bonhoeffer-van der Pol or Morris-Lecar; 3D models of spiking/bursting (Hindmarsh-Rose), high-dimensional biophysically motivated models (Hodgkin-Huxley), multi-compartment models, distributed-parameter models, and many others. Connections between such individual neurons include simple coupling, excitatory, and inhibitory synaptic connections, etc.

We refer readers to Ref.\cite{GerstnerBook} for the overview of the possible systems mentioned above. Herein, however, we will consider two particularly popular neuronal models ~\cite{Bonhoeffer48,Hindmarsh-Rose-84} with the sole goal of mimicking various realistic signalling patterns of collective neuronal activity \emph{qualitatively}: namely, regular, chaotic, and bursting signalling regimes. 

\emph{Bonhoeffer--van der Pol oscillators.} 
As our first basic model, we consider a population of $N$ regularly oscillating neurons, known as Bonhoeffer--van der Pol or FitzHugh--Nagumo oscillators, globally coupled via the mean field X. See Fig.~\ref{fig:TSall}(a) for an illustration of its oscillatory behavior (for $t < 5000$). The equations governing the model are:
\begin{equation}
\begin{cases}
\dot{x}_k &= x_k-\frac{x_k^3}{3} - y_k +I_k +\e X +A\;,\\
\dot{y}_k &= 0.1 (x_k-0.8y_k+0.7)\;,\\
\end{cases}
\label{eq:bvdp}
\end{equation}
where $k=1,\ldots,N$ is the index of the neuron, where $X=\frac{1}{N}\sum_k x_k$ is the mean field, and where $A$ is the action.
The neurons are not identical: the currents $I_k$ are drawn from a Gaussian distribution with a mean of $0.6$ and a standard deviation of $0.1$. The strength of the global coupling is determined by $\e$.

\begin{figure}[h!]
\centering
\includegraphics[width=1\columnwidth, left]{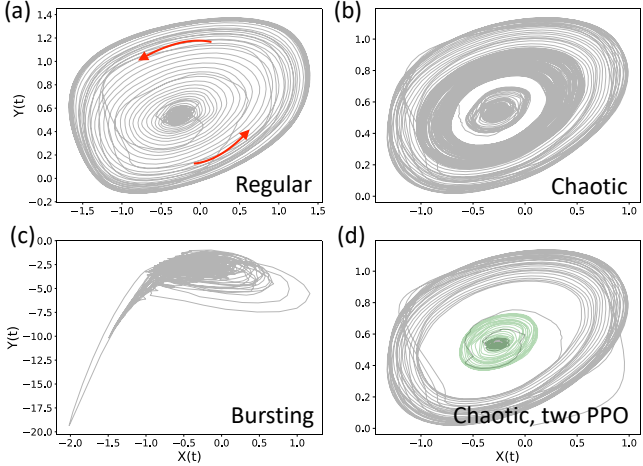}
\caption{Phase portraits of suppression dynamics. Arrows indicate a direction along which trajectories gradually reach ``special point'' $\{X_0,Y_0\}$ where the ensemble does not have the collective mode. The green part of the trajectory in \textbf{(d)} belongs to the secondary PPO, trained to suppress chaotic weak-amplitude oscillations.}
\label{fig:TSphase}
\end{figure}

This model has two properties that make the control problem non-trivial. First, for very low values of the coupling $\e$, the mean fields are $X_0\approx -0.27$, $Y_0\approx 0.55$,
i.e. the fixed point to which the system should converge is not the origin and is {\it a priori} unknown.
Second, the model exhibits chaotic collective dynamics for certain values of $\e$ (Chaotic model, see the broadened trajectory in Fig.~\ref{fig:TSphase}(b)).

\emph{Bursting Hindmarsh--Rose neuronal model.}
The other type of oscillators considered is an ensemble of Hindmarsh-Rose \cite{Hindmarsh-Rose-84} neurons in a bursting regime:
\begin{equation}\label{eq:HRn}
\begin{cases}
 {\dot x}_k &= 3x_k^2 - x_k^3 + y_k - z_k + I_k +\e X +A\;,  \\ 
 {\dot y}_k &= 1 - 5x_k^2 - y_k \;, \\ 
 {\dot z}_k &= 0.006[4 (x_k +1.56 ) - z_k]\;. \\ 
 \end{cases}
\end{equation}
The currents $I_k$ are also drawn from a Gaussian distribution with mean  
$3$ and standard deviation $0.02$. For illustration, see the Bursting model in Figs~\ref{fig:TSphase}(c) and \ref{fig:TSall}(c).

The collective dynamics of both systems (\ref{eq:bvdp}) and (\ref{eq:HRn}) are illustrated by the phase portraits shown in Fig.~\ref{fig:TSphase}, 
where we plot $Y=\frac{1}{N}\sum_k y_k$ vs. $X$ for different values of the coupling strength $\e$ (Regular corresponds to $\e = 0.03$ and Chaotic to $\e = 0.02$ in Eq. (\ref{eq:bvdp}); Bursting pattern is $\e = 0.2$ in Eq. (\ref{eq:HRn})). 

\subsection{Action and State}
The action and the state are respectively the input to and the resulting response output from the environment, produced with a sampling rate $\Delta$.

We consider idealistic $\delta$-shaped pulse actions, with a constant interval $\Delta$ between each pulse and an amplitude limited by a value $A_{max}$. The action $A(t_n)$ is tuned at each time step, with $-A_{max}\leq A(t_n)\leq A_{max}$ and $t_n=n\Delta$, $n=1,2,\cdots$. 
For treatment of more realistic pulses, encountered in the DBS systems, see~\cite{krylov2019RLChaos}. For convenience, below we omit the index $n$ for the discrete time $t_n$. Naturally, smaller values of $A(t)$ are commonly sought after in biological applications, such as DBS for Parkinson's disease, as the system should be as little invasive as possible. The total ``energy'' supplied to the ensemble from an external source, $A_\text{total}=\sum_t A(t)$, is thus another measure that one aims to minimize in practice. The action affects all neurons similarly, its precise effect is represented by the letter A in Eqns.~(\ref{eq:bvdp}) and (\ref{eq:HRn}).


The state is based on the current value of the mean field, $X(t)$, extracted using a Runge--Kutta-based solver for Eqns.~(\ref{eq:bvdp}) or (\ref{eq:HRn}). The solver is implemented in the gym environment we developed. This provides feedback from the system, after application of action $A(t)$. To account for the oscillatory behavior of the model, the state $X_{\text{state}}$ consists of the $M=250$ most recent values of $X$.

Some of our experiments will introduce some noise in the action: the executed action is the one selected by the agent plus a white noise term. Similarly, to mimic real-world conditions, we will also introduce some noise at the state perception level.

\subsection{Reward}
For a given action $A$ and a given observation $X_{\text{state}}$ at time $t$, we propose the following class of reward functions for synchrony suppression tasks: 
\begin{equation}\label{eq:reward-general}
R\big[t\big] =-\big(X(t)-\langle X_{\text{state}}\rangle_t\big)^2 - \beta|A(t)|,
\end{equation}
where the first term rewards convergence of the system to an average of the mean field over previous $M$ values, $\langle X_{\text{state}}\rangle_t=M^{-1}\sum_{l=1}^M X(t-l+1)$, and the second term favors smaller values of the action $A$.
The coefficient $\beta$ allows to introduce a bias towards a desired outcome (\textit{e.g.}, a more accurate convergence to a particular value of the mean field $X$ \emph{vs.} a smaller amplitude of the suppression pulse).

\subsection{Agent}
We trained our RL agent using the Proximal Policy Optimization algorithm~\cite[PPO]{RL-PPO-schulman2017}. We briefly describe the method below. As usual in RL, we wish to maximize the expected return, defined as the discounted sum of rewards: 
 \begin{equation}\mathcal{R}_\pi(\theta) =\mathbb{E}_{\pi}\Big[\sum^{\infty}_{t=0}\gamma^{t}\hspace{1pt}R\big[t\big]\Big],
 \label{eq:reward-Total}
\end{equation}
where $\mathbb{E}_{\pi}$ is the expectation over visited states following a given policy $\pi$, $\gamma$ is a discount factor that controls the trade-off between long-term and immediate rewards (set to 0.99 in our experiments), and $R\big[t\big]$ is the reward received at time $t$, specified by Eq.~(\ref{eq:reward-general}).

The policy is parameterized by a neural network with parameters $\theta$, encoding the probability of taking action $A$ when the current state $X_{\text{state}}$ is $X$:
\begin{equation}
\pi = \pi_\theta(A | X)=\mathbb{P}_\theta\big\{A(t)=A\big|X_{\text{state}}=X\big\}.
\label{eq:pi}
\end{equation}

\begin{figure}[h!]
\centering
\includegraphics[width=1\columnwidth, left]{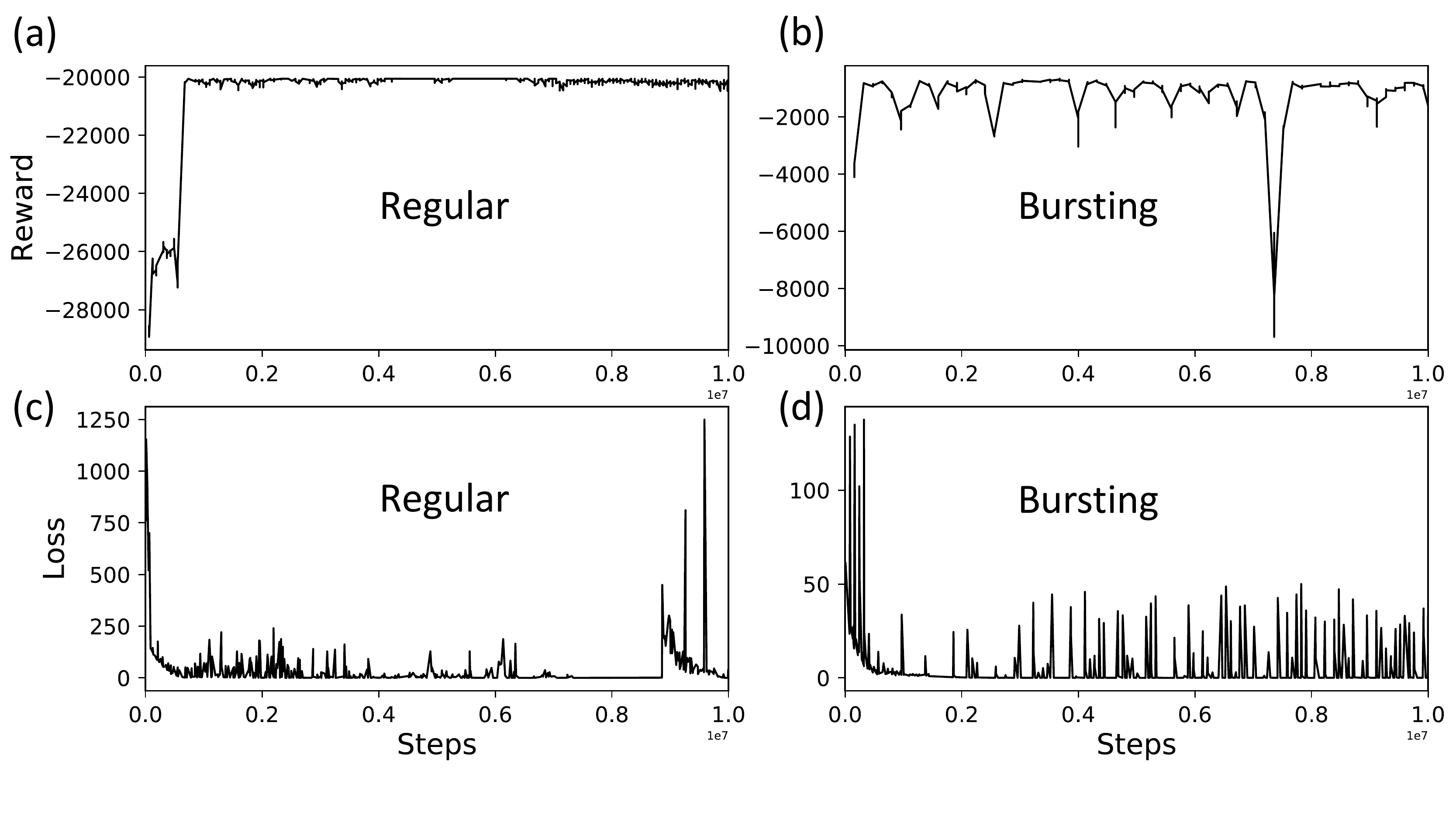}
\caption{Reward and Loss curves for ensembles of $N$=1000 neurons evolving according to the regular and bursting models.}
\label{fig:rew}
\end{figure}

$\theta$ is optimized using PPO to maximize the expected return given by Eq.~(\ref{eq:reward-Total}). In our experiments, we used two-hidden layers MLPs with 64 neurons, trained using the Stable Baselines library~\cite{RL-stable-baselines}, with the default parameters for PPO. Generally speaking, the nonlinear nature of Eqns.~(\ref{eq:bvdp}) and (\ref{eq:HRn}) will make the feedback highly sensitive to the amplitude of the input. To handle this sensitivity, we opted for the use of two agents trained for different values of neuronal spiking activities. Given the small size of the networks, training was performed on CPU~\footnote{The code is available at {https://github.com/cviaai/RL-DBS/}}. The training reward (Eq.~\ref{eq:reward-Total}) and PPO loss are plotted for the Regular and Bursting environments in Figure~\ref{fig:rew}.

\section{Results}

\subsection{Synchrony suppression in the environments}
\begin{figure}[h!]
\centering
\includegraphics[width=1\columnwidth, left]{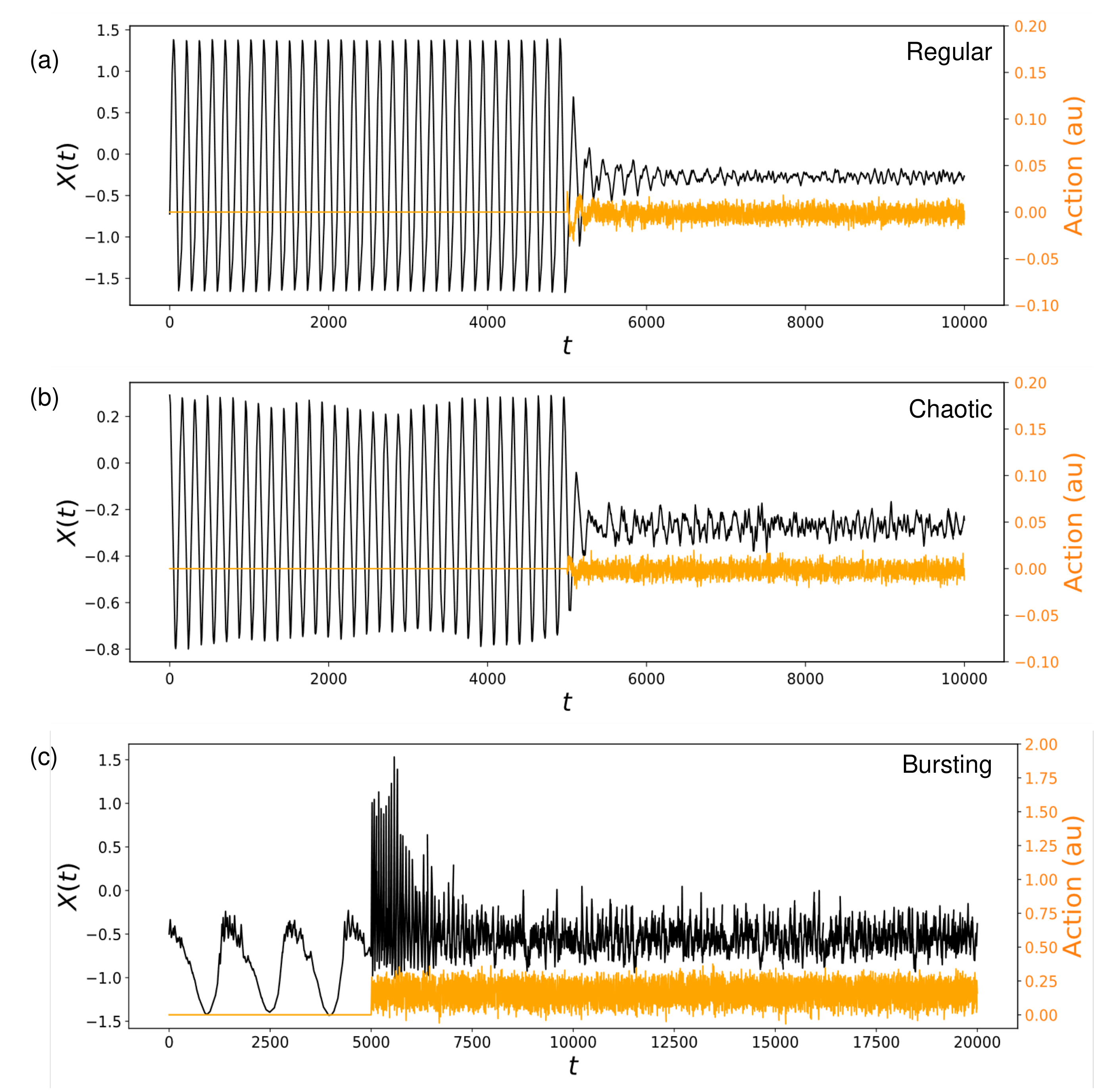}
\caption{Suppression of synchrony in a population of $N=1000$ neurons described by \textbf{(a)} Bonhoeffer-van der Pol model with $\e = 0.03$ (regular regime), \textbf{(b)} same, with coupling strength $\e = 0.02$ (chaotic regime), \textbf{(c)} Hindmarsh-Rose model (bursting regime with coupling $\e=0.2$). Are plotted the mean field (top black curve) and action pulses used for suppression (bottom orange curve, plotted against the right axis in the same units as the mean field).}
\label{fig:TSall}
\end{figure}

We first test our agent on an ensemble of $N=1000$ self-sustained Bonhoeffer-van der Pol neurons oscillating around a non-zero equilibrium point and globally coupled with $\e = 0.03$ (Fig.~\ref{fig:TSall}(a)). At $t=5000$, we initiate synchrony suppression by sending action pulses according to our trained PPO agent. This confirms that the reward function described by Eq.~(\ref{eq:reward-general}) for $\beta=2$ leads to convergence to the natural equilibrium point, with a non-zero average $X_0\approx -0.2669 \pm 0.0016$. At $t=5000$, \textit{i.e.} when suppression is activated, the action amplitudes spike slightly, for about 200 time steps, and then quickly reduce to $\sim 0.01$. As a point of comparison, we study below the impact of constant actions. We wish to emphasize that each individual neuron maintains its output; it is the desynchronization of the entire ensemble that causes the mean field to decrease.

RL can also suppress synchronization in the Bonhoeffer-van der Pol ensemble when the collective mode is chaotic ($\e = 0.02$, Fig.~\ref{fig:TSall}(b)). Although the oscillatory dynamics is now irregular, our PPO agent performs here similarly to the non-chaotic regimes, with $X_0\approx -0.2707 \pm 0.0018$, the same order of magnitude for the required action amplitudes ($\sim 0.01$), and a total stimuli energy $A_\text{total}$ required for suppression only $8\%$ larger than in the regular regime.

The bursting output of Hindmarsh-Rose neurons, Eq.(\ref{eq:HRn}) for $\e=0.2$ and $N=1000$ can also be suppressed (Fig.~\ref{fig:TSall}(c)). The bursting pattern and the high synchrony of the oscillators occurs at the beginning until a series of action pulses is applied at $t=5000$. 
Interestingly, immediately after the stimuli are applied, the mean field spikes above its anterior value, which portrays a transient regime where the system undergoes a temporal increase of synchrony.
As the PPO agent continues to adapt to the current state, the synchrony of oscillations vanishes, at which point (around $t=6100$) the mean field converges to the special point $X_0=-0.5308 \pm 0.0659$.

The convergence of the ensemble to the special point $X_0$ is best monitored in the phase space $\left\{X(t),Y(t)\right\}$, shown in Fig.~\ref{fig:TSphase}(c). As the agent acts on the collective oscillation, the trajectories bend towards the fixed point. Broadening of the trajectory in the chaotic and in the bursting ensembles have particular signatures indicating intricate signalling regimes.

\subsection{Multiple PPO agents}
Dynamical nonlinear systems containing large populations of coupled neurons are especially hard to control because of their very different responses to weak and strong stimuli. This is where another modern direction of RL, entailing multiple agents, could be beneficial for the task at hand. We propose to use multiple \emph{auxiliary} PPO agents, trained on various neuronal patterns, \textit{e.g.} during the transient ones occurring immediately after $t=t_\text{on}$ ($5000$ in our experiments) or during the suppressed regime. As such, the primary agent would ``see'' only the strong stimuli, whereas the auxiliary agent would ``see'' only the signal that has already been partially suppressed and is, therefore, weaker.  Figure 5 demonstrates that when this secondary model overtakes the control at $t=10000$, it further reduces the amplitude of the mean field $X$ and desynchronizes the ensemble beyond the performance of a single model.

Indeed, the response to a stimulus is determined by the corresponding phase response curve that does not depend on the stimuli amplitude only in the limit of an infinitely small action~\cite{Canavier:2006}; with a finite action, the response will always be pronounced as dependent on the amplitude of the input. Long-term, one could envision a library of such ANNs pre-trained at different amplitude levels, at different values of sampling rate $\Delta$, and at different pulse skipping rates $\kappa$ -- all to be embedded into the software controlling a DBS device. This promises a personalized approach to the patients with different signaling patterns and at different progression stages of the disease, regardless of its etiology. Characterization of the full nonlinear response of these strongly interconnected ensembles and engaging three or more such agents will be studied in future work. Deep architectures, alternatively, are also expected to fit the nonlinear response curve better than the small networks we used in our study, albeit with the associated lack of physical interpretability.

\begin{figure}[h!]
\centering
\includegraphics[width=1\columnwidth]{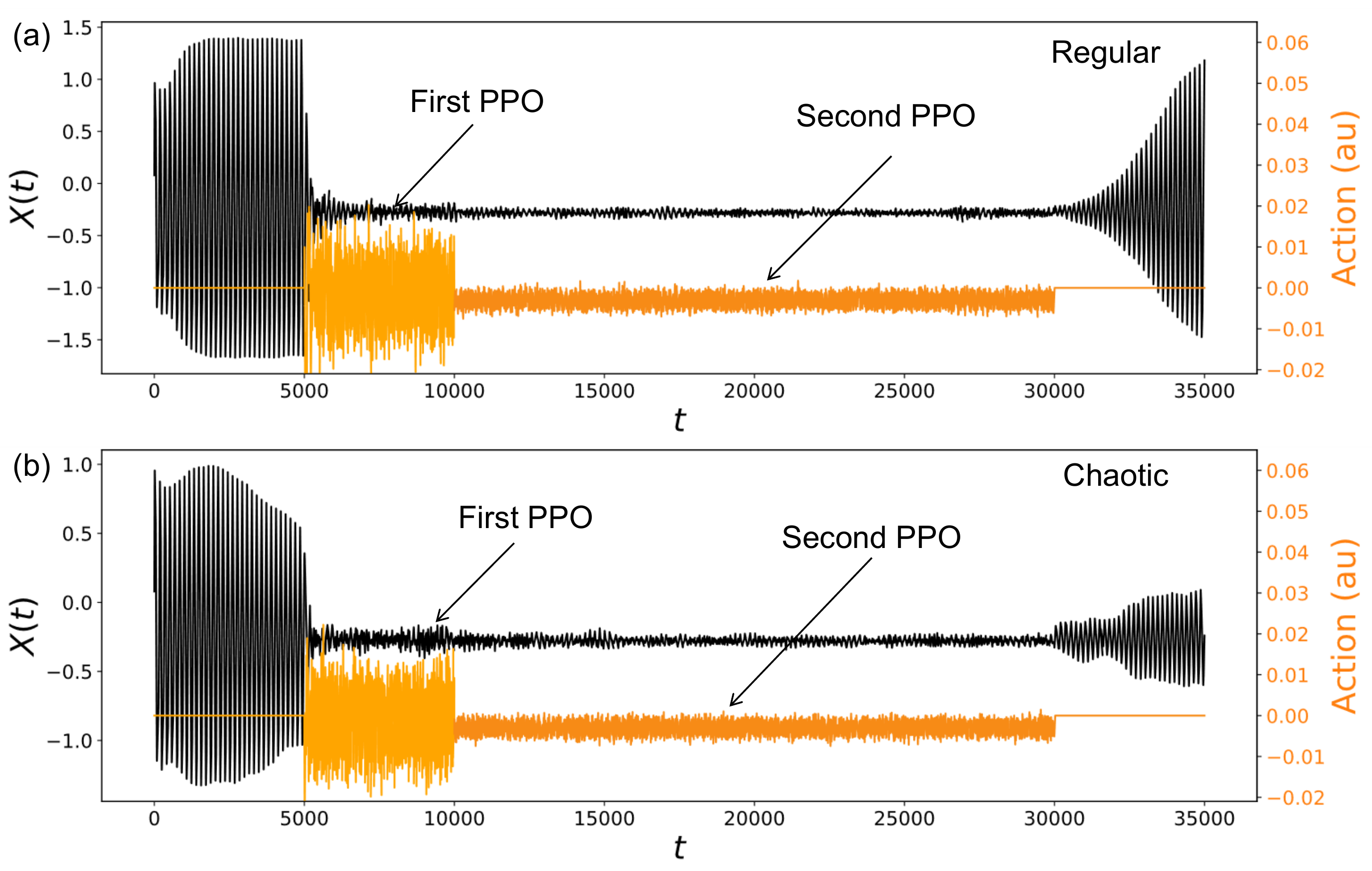}
\caption{Demonstration of suppression on the regular model using two PPO agents. The first one suppresses activity with strong mean field amplitude, whereas the second is activated when the initial synchrony is already sufficiently removed (arrows show the corresponding activations for the regular \textbf{(a)} and chaotic \textbf{(b)} models). Notice the reduced variance of the mean field and the smaller amplitude of the stimuli after engaging the second agent.}
\label{fig:Regular-TS-2ANNs}
\end{figure}

\subsection{Quantitative analysis}
We now proceed to characterize the RL-based suppression as a function of various parameters of the system and of the stimulation.
The major factor that determines the amplitude of the collective oscillation is the coupling strength $\e$, which we thoroughly varied. The results for the Bonhoeffer--van der Pol ensemble, Eq.~(\ref{eq:bvdp}), are shown in Fig.~\ref{fig:sup-coef}. 
For the unperturbed system, the dependence of the standard deviation of the collective mode, std$(X)$, on the coupling strength $\e$ follows a threshold-like curve, Fig.~\ref{fig:sup-coef}(a).
The value for std$(X)$ was taken when the PPO agent reached the best possible level of synchrony suppression. As can be seen in Fig.~\ref{fig:TSall}(b), 
this final ``steady stage'' of the control is achieved soon after the stimuli application is switched on, at about $t_\text{steady}=5200$, and is preserved until the control is switched off. The corresponding value for the Hindmarsh-Rose model is $t_\text{steady}\approx 7000$, see Fig.~\ref{fig:TSall}(c).

In the suppressed steady state the mean field continues to fluctuate due to the final size of the ensemble ($\frac{1}{\sqrt{N}}$~\cite{Pikovsky-Ruffo-99}). The amplitude of the action also fluctuates but the pulse sequence now has a uniform variance and a diminished range of amplitudes required to keep the control active. We speculate that there is an additional source of fluctuations emerging from the probabilistic uncertainty inherent to the ANNs. Despite not reaching the theoretical limit, the RL algorithm is actually more pertinent to the real experimental data because this uncertainty can indirectly train the model to accommodate noisy signals.

The extent of the mean field suppression can be quantified by the following suppression coefficient
\begin{equation}
  S=\frac{\text{std}\big[X_{\text{before}}\big]}{\text{std}\big[X_{\text{after}}\big]}.
  \label{eq:supp}
\end{equation}
where $X_{\text{before}}$ (resp. $X_{\text{after}}$) represents the mean field values before (resp. after) the stimuli application. The fluctuations of the suppressed field do not depend much on $\e$, but the amplitude of the collective mode of the unperturbed field grows with $\e$, see Fig.~\ref{fig:sup-coef}(a). The suppression coefficient is maximal for strongly synchronized systems and achieves $S \approx 33$ in that case.

\subsubsection{Study of skipping pulses}
\begin{figure}[h!]
\centering
\includegraphics[width=1.1\columnwidth]{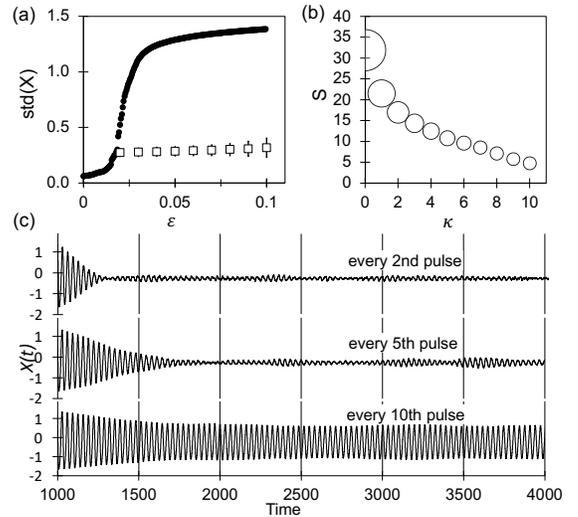}
\caption{Quantitative analysis of suppression via RL. \textbf{(a)} Std of the mean field $X$ \emph{vs} coupling strength $\e$. Dots show dependence before suppression and boxes show the std values after the transient period, when a steady suppressed state is achieved. Error bars are standard deviations calculated over 10 experiments. \textbf{(b)} Suppression coefficient as a function of $\kappa$ (the skip parameter). Bubble sizes are proportional to total supplied energy $A_\text{total}$. \textbf{(c)} Suppression for a PPO agent trained to suppress oscillations every time step, but allowed to interact with the environment every $\kappa^{\text{th}}$ time step.}
\label{fig:sup-coef}
\end{figure}

Next, of great importance for future RL-based DBS devices, is the minimization of total energy sent via stimuli to the brain. We analyzed the dependence of $S$ on a skip parameter $\kappa$, defined as follows. We trained a PPO agent as though to send a stimuli every time step $\Delta$, but only allowed it to send pulses to the environment every $\kappa^{\text{th}}$ time steps.
The rationale behind this test is to look for the optimal frequency of action pulses in order to minimize the energy of the perturbation sent to the system while still suppressing synchrony.
The resulting fall in the suppression efficiency shown in Fig.~\ref{fig:sup-coef}(b) can be deemed as a classic example of trade-off when \textit{e.g.} a limited stimuli energy $A_\text{total}$ must be used or an incomplete suppression is desired. Figure~\ref{fig:TSall}(c) shows the time dependence of the mean field immediately after the stimuli are initiated at $t_{\text{on}}=1000$ for the case of $\e=0.03$ and $N=1000$ and for different values of $\kappa$. As we can see, for $\kappa = 5$, suppression is still rather efficient and comes with a smaller total energy supplied to the system (circle diameters in Fig.~\ref{fig:sup-coef}(b)).

\subsubsection{Study of response to constant stimuli}

A standard test of an RL environment is to explore the efficiency of constant stimulation (here, de-synchronization is no longer the task). To study such a response we use our simulator for the Bonhoeffer--van der Pol model and predict the evolution of $X(t_n)$ for constant values of $A(t)$ ranging from $-0.1$ to $0.1$ with a step size of $0.01$. Outside that range, the effects are simply more pronounced, and less desirable. Results are shown in Fig.~\ref{fig:noise}(a). For relatively negative values of the action, we do observe suppression of the oscillations, which implies that the very individual neurons cease to oscillate. In these stable cases, the mean value of $X$ is around $-1$ and the applied pulses are larger than $0.06$ in absolute value. In contrast, using our trained agent, we achieve the same level of suppression with a mean $X$ of $-0.26$ and an average action smaller than $0.010$ in absolute values (with a standard deviation of $0.002$): the RL agent is far less invasive, and sends far less energy to the system. Finally, for constant actions that are smaller than $0.06$ in absolute value, we see that suppression is very limited.

\subsubsection{Study of Action-State noise stability}
\begin{figure}[h!]
\centering
\includegraphics[width=1\columnwidth]{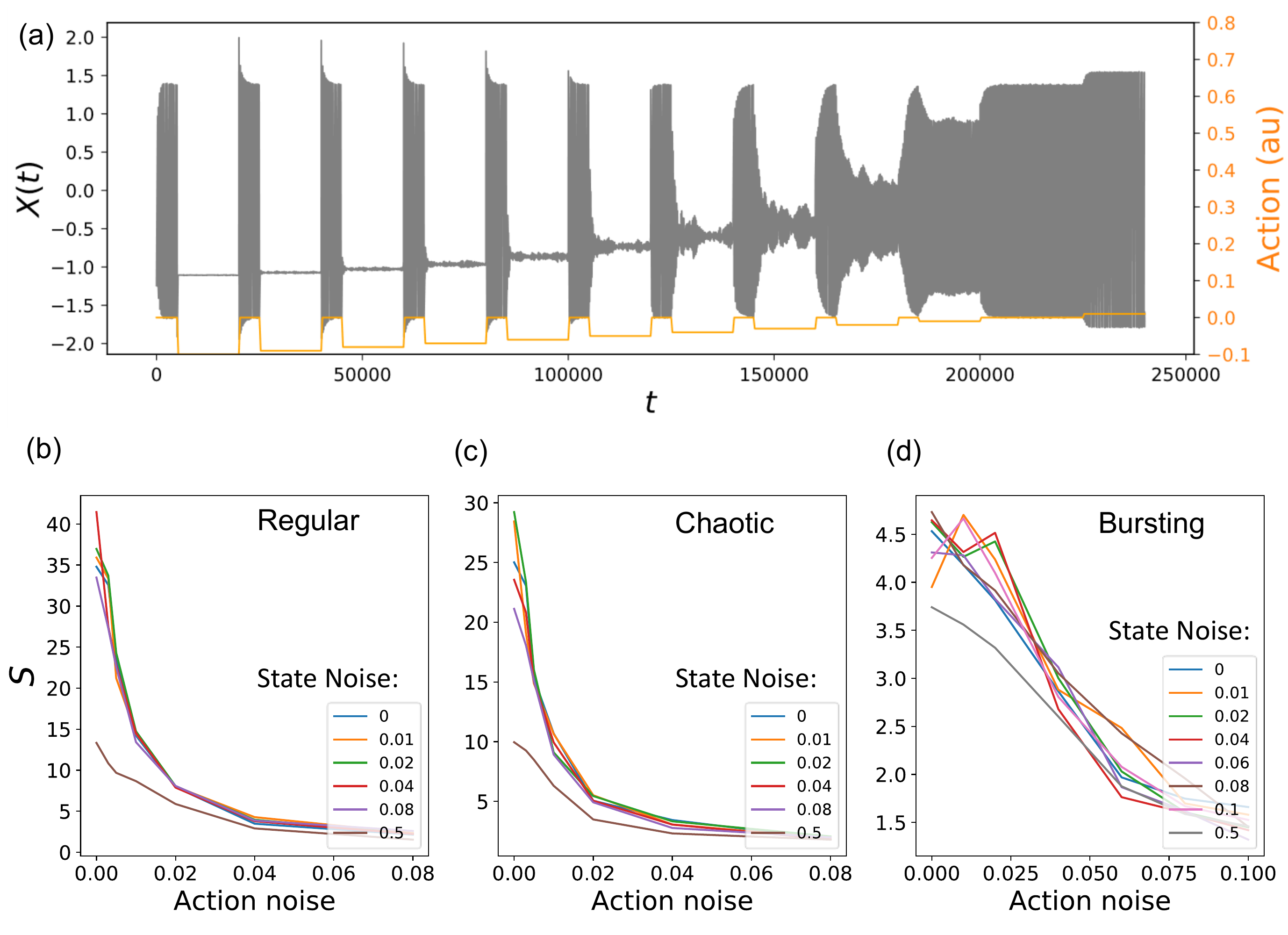}
\caption{Characterization of model's stability to (a) constant stimulation and to (c-d) Action-State noise for three oscillatory regimes.}
\label{fig:noise}
\end{figure}

Another essential condition for the deployment of RL agents to real-world scenarios is their stability to noise. Observations of the actual state will never be accurate, nor will the stimuli applied to the brain be exactly the one required by the agent. For these reasons, we ran suppression experiments in a noisy setting. For the three types of environments (regular Bonhoeffer--van der Pol, chaotic Bonhoeffer--van der Pol, and bursting Hindmarsh--Rose), we added some white noise to the state $X(t)$ observed by the RL agent (at each time step, drawn independently from $\mathcal{N}(0,\sigma_x^2)$). Similarly, the action performed in the environment was the action selected by the agent with some additive noise (drawn from $\mathcal{N}(0,\sigma_a^2)$. Fig.~\ref{fig:noise} shows the suppression coefficient at the end of training for various values of $\sigma_x$ and $\sigma_a$, each point corresponding to an average over $5$ seeds. We first observe that the state noise has a limited effect on the efficiency of the trained agent.

Noisy actions have a far more significant impact on the efficiency of the agent. For the Bonhoeffer--van der Pol environments considered in this section, the mean action is approximately $-0.002$ (depending on $\varepsilon$ and the randomness of the run). Applying a noise of the same order of magnitude, the agent reaches a similar suppression coefficient $S \approx 25$. For larger noise levels, we observe a steady degradation in performance. 

The same conclusions can be drawn for the bursting model, although noise levels need to be larger to observe a decreased performance. Indeed, in the Hindmarsh--Rose environment, the mean action is $\approx -0.2$ with a standard deviation of $0.05$, it is thus more robust to small perturbations of the applied stimuli. Overall, these experiments allow the definitions of thresholds below which stability to noise is guaranteed.

\section{Discussion and State-of-the-art}

The speed of suppression and the residual synchrony in the time series curves in Fig.~\ref{fig:sup-coef}(c) portray the trade-off between supplied energy and the extent of residual synchronization mentioned above. The fact that the five-fold reduction of the stimuli frequency still allows achieving a satisfactory degree of suppression naturally suggests the following pathway for future work. We speculate that the most efficient application of stimuli should actually be non-uniform pulse trains in time and that the frequency of it should be adapted according to the patient's symptoms. 

However, as mentioned above, the cause-effect relationship between the synchrony and the pathology is still an unproven hypothesis in neurobiology and in computational neuroscience. Nonetheless, machine learning methods could be proposed for the optimization of the stimulation parameters regardless of the etiology of the disease, and -- as we studied on the synthetic data -- RL could be considered as the ideal candidate for integration with a real DBS device. Pre-clinical approbation~\cite{ParkinsonAnimalModel} could be a logical continuation to test both the cause-effect hypothesis and to optimize device settings experimentally prior to proceeding to human studies.

But perhaps more importantly, the community needs to standardize and honestly compare various control algorithms apple-to-apple - something that is not possible to accomplish as of today. As of now, the schemes proposed in the literature exploited delayed or non-delayed, linear and nonlinear control loops, continuous or pulsatile stimulation, specialized pulses that preserve total charge, adaptive tuning of the feedback parameters. And recently, ML-based approaches started to appear. Having different input parameters, different underlying models' assumptions, and different criteria to define successful suppression, the current state of affairs suggests that our gym environment holds the potential to become particularly useful and to provide a unified platform to evaluate various methods.

In our work, we supply a potentially large and diverse collection of RL environments within a single framework. Pulsatile, continuous, or purposefully optimized agents could interact with these environments effectively enabling the parameter search for a particular configuration of a DBS device. 

Having introduced a clear metric (Eq.~\ref{eq:supp} and that of a total supplied energy $A_\text{total}=\sum_t A(t)$) as a criterion for efficient suppression, and having characterized basic collective behavior seen in neuronal ensembles (regular, chaotic, bursting), we aspire to enable a ``gym research'' effort that is easy to set up and use. The proposed framework should make it easy to reproduce published results across physics and computer science publications and to compare results from different papers. Clear metrics and synthetic data can also become a sound platform for various AI competitions.

\section{Conclusions}
To conclude, we presented a new RL gym framework for the synchrony suppression task in a strongly interconnected oscillatory network that is believed to be the cause of tremor and other systemic neurological symptoms. Considering limit-cycle Bonhoeffer-van der Pol oscillators and Hindmarsh-Rose neurons as the test models, we demonstrated successful synchrony suppression for regular, chaotic, and bursting collective oscillation, without having knowledge about the ensemble model.

An important advantage of the RL-based suppression method is that it is data-driven and universal. It could be readily implemented in an experimental setting if one takes the measuring/stimulating equipment characteristics and limitations into account. The suppression workflow proposed in the diagram of Fig.~\ref{fig:diagram} is universal and can be exploited for a variety of practical tasks. We find Reinforced Learning to be an ideal candidate for clinical approbation as a ``smart'' control algorithm to be embedded into deep brain stimulation devices.
\bibliographystyle{named}
\bibliography{main}

\end{document}